# Forecasting the Impact of Connected and Automated Vehicles on Energy Use: A Microeconomic Study of Induced Travel and Energy Rebound

Morteza Taiebat[†,‡], Samuel Stolper[†], Ming Xu[†,‡]

[†] School for Environment and Sustainability, University of Michigan, Ann Arbor, MI, USA
[‡] Department of Civil and Environmental Engineering, University of Michigan, Ann Arbor, MI, USA
{taiebat, sstolper, mingxu}@umich.edu

**Abstract:** Connected and automated vehicles (CAVs) are expected to yield significant improvements in safety, energy efficiency, and time utilization. However, their net effect on energy and environmental outcomes is unclear. Higher fuel economy reduces the energy required per mile of travel, but it also reduces the fuel cost of travel, incentivizing more travel and causing an energy "rebound effect." Moreover, CAVs are predicted to vastly reduce the *time* cost of travel, inducing further increases in travel and energy use. In this paper, we forecast the induced travel and rebound from CAVs using data on existing travel behavior. We develop a microeconomic model of vehicle miles traveled (VMT) choice under income and time constraints; then we use it to estimate elasticities of VMT demand with respect to fuel and time costs, with fuel cost data from the 2017 United States National Household Travel Survey (NHTS) and wage-derived predictions of travel time cost. Our central estimate of the combined price elasticity of VMT demand is -0.4, which differs substantially from previous estimates. We also find evidence that wealthier households have more elastic demand, and that households at all income levels are more sensitive to time costs than to fuel costs. We use our estimated elasticities to simulate VMT and energy use impacts of full, private CAV adoption under a range of possible changes to the fuel and time costs of travel. We forecast a 2-47% increase in travel demand for an average household. Our results indicate that backfire – i.e., a net rise in energy use – is a possibility, especially in higher income groups. This presents a stiff challenge to policy goals for reductions in not only energy use but also traffic congestion and local and global air pollution, as CAV use increases.

**Highlights:**

- We develop a microeconomic model of VMT choice under time and budget constraints.
- Using NHTS data, we estimate VMT elasticities with respect to fuel and time costs.
- We use these elasticities to forecast CAV-induced travel and energy rebound.
- A 38% drop in time cost offsets energy savings from a 20% fuel efficiency rise.

**Graphical Abstract:**

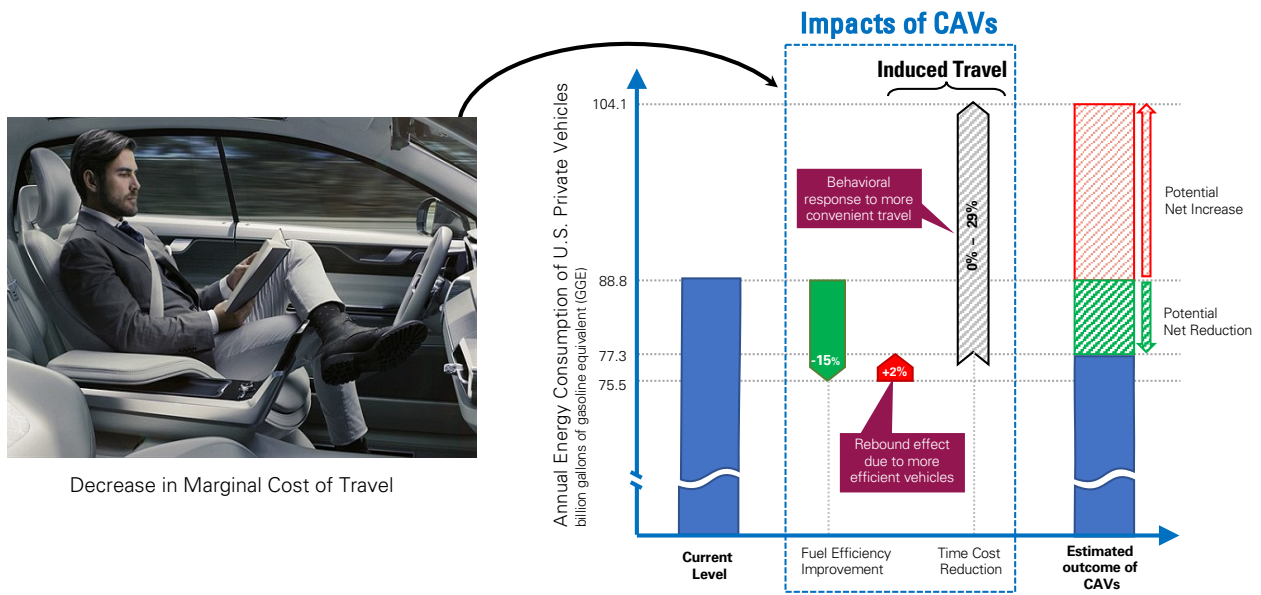

**How to cite:** Taiebat, M., Stolper, S. and Xu, M., 2019. Forecasting the Impact of Connected and Automated Vehicles on Energy Use: A Microeconomic Study of Induced Travel and Energy Rebound. _Applied Energy_ 247, 297-308.

# 1. Introduction

Connected and automated vehicle (CAV) technology is expected to be an indispensable but disruptive factor in the transportation sector, transforming the mobility paradigm, transportation markets, and travelers' behavior in the coming decades. It will likely increase transportation safety to an unprecedented level [1], enhance mobility, provide a higher level of comfort and convenience for travelers, and reduce the cost of driving for individuals, all of which will be welfare-improving for society. At the same time, vehicle connectivity and automation will inevitably and significantly change energy demand in the transportation sector. The extent of these changes is still largely unclear [2–4] and yet will have major consequences for energy supply and the environment alike.

Several characteristics of CAV[1] technology will influence energy consumption, including improvements in route optimization, eco-driving, crash avoidance, and vehicle right-sizing, among others [2]. Many of these improvements will push energy use downwards; however, some will very likely work in the opposing direction. Chief among the factors that will exert upward pressure on energy demand is the marginal cost of driving, which is expected to drop significantly with CAV technology. Higher fuel economy of CAVs [2,5,6] will cause the per-mile fuel cost of travel to drop. This, in turn, will induce additional travel that partially offsets the fuel savings of energy efficiency – commonly referred to as a "rebound effect"[2]. In addition, increased comfort and reduced attention requirements[3] will cause the per-mile *travel time cost* to drop [7], inducing even more additional travel [2,5,8,9].

The key parameter dictating the magnitude of travel demand induced through these channels is the elasticity of travel demand with respect to the price of travel [10–13]. The overwhelming majority of existing studies on the energy impact of more efficient vehicle technologies focus exclusively on the fuel-cost component of the price of travel [14–20]. Consequently, such studies are unlikely to have external validity in the context of vehicle automation, which will intimately affect both fuel cost *and* time cost. While recent research on the energy use impacts of vehicle automation does consider the impact of time cost changes (for example, Wadud et al. [5]), it tends to borrow fuel and time cost elasticities that are estimated elsewhere, in isolation from each other, and without the aim of developing CAV-specific predictions. Most studies focus on how changes in mobility – especially changes in the vehicle-level energy efficiency of CAVs – affect energy use, *holding travel demand constant* (for instance, [21–24]). The assumption of fixed demand almost certainly leads to overestimation of the environmental benefits of this technology [2].

---

[1] CAVs are also referred to as "autonomous", "self-driving", or "driverless" vehicles interchangeably in the literature, though these are not the same. For a disambiguation of definitions, refer to [2].

[2] The rebound effect can refer to the general phenomenon of increased driving after a rise in fuel economy, or it can be mathematically defined as the percent change in miles traveled caused by a one-percent change in fuel economy (or, relatedly, a one-percent change in fuel costs). The empirical investigation of micro-level rebound usually utilizes regression-based approaches with cross-sectional, time series, or panel data [14,19].

[3] This is viewed as a likely feature of high levels of automation (level 3 and above) [1].

In this paper, we use the most recent empirical microdata available to estimate the elasticity of travel demand with respect to the marginal fuel and time costs of travel in a single, unified framework. Our approach adapts standard microeconomic modeling and statistical techniques to account for the value of time in elasticity estimation. We first specify a theoretical model of consumer utility maximization from vehicle-miles traveled (VMT) and other goods, subject to time and income constraints. The model illustrates how the opportunity cost of time spent traveling and the fuel cost of travel affect the privately-optimal choice of VMT. From it, we derive an estimating equation for the combined, fuel- and time-inclusive price elasticity of VMT. We fit several specifications of this equation using household-level vehicle and travel data from the 2017 United States (U.S.) National Household Travel Survey (NHTS) [25] as well as predictions of travel time cost based on reported income. The resulting empirically-derived elasticity estimates allow us to forecast the changes in travel demand induced by CAV technology, as well as the associated energy rebound effects.

Our study produces three key findings. First, our central estimate of the combined, fuel- and time-inclusive price elasticity of demand for VMT is -0.39. This is significantly larger than the

-0.06 to -0.28 range found in existing studies of the fuel price elasticity of demand [17–20] and significantly smaller than the -1.0 to -2.3 range found in studies of demand elasticity with respect to the generalized cost of travel[4], the latter of which is cited in prior work on CAV-induced travel demand [4,5]. Replicating our procedure with 2009 NHTS data yields a similar central estimate of -0.45. Our results highlight the importance of accounting for the opportunity cost of time in travel demand elasticity estimation and suggest that existing predictions of CAV-induced travel may not be based on relevant travel demand parameter values.

Second, travel demand elasticities exhibit significant heterogeneity that inform future forecasting methodology and policy discussions. We find that households respond very differently, on average, to fuel price changes versus time cost changes. Our preferred estimate of the fuel price elasticity is -0.1, while our preferred estimate of the time cost elasticity is -0.4. Moreover, all of our elasticity estimates vary significantly with income. We find that richer households have less elastic demand with respect to fuel costs but more elastic demand with respect to time costs. The aggregate, fuel- and time-inclusive price elasticity of VMT rises with income; for example, the average elasticity of the upper three groups is 64% larger than that of the bottom group. In other words, our estimated model predicts that relatively richer households will increase their travel relatively more in response to automation and thus stand to experience greater welfare gains.

Third, the aggregate, CAV-induced reduction in energy use may be quite small or even negative. In our model, the magnitude of this reduction depends on (a) elasticities of demand with respect to the price of travel, (b) projected increases in fuel economy of CAVs, and (c) projected decreases in travel time cost with CAVs. We use our estimates of (a) to simulate induced VMT for different combinations of (b) and (c). The range of possible impacts of CAVs on VMT, and thus energy consumption, is wide. However, backfire – a net rise in energy

---

[4] In transportation economics, "generalized cost" refers to the sum of monetary and non-monetary costs of a trip. For instance, the generalized cost of private vehicle travel includes total cost of ownership (TCO, including capital, fixed, and operation costs) and monetized passenger travel time [13].

consumption – is a distinct possibility, because high-income households have large elasticities of demand and also high baseline energy use. This, in turn, implies the possibility of net rises in local and global air pollution.

Ultimately, the energy and environmental impacts of CAV technology will depend on not just changes in the marginal cost of travel, but also the capital cost of an automated vehicle, the safety benefits of automation, and changes in ride- and vehicle-sharing, among other aspects of the mobility transition. The very non-marginal nature of the upcoming mobility transition presents steep challenges to researchers who seek to provide rigorous predictions of future travel behavior and energy use. Our contribution is to use the most recent microdata available in the United States to develop empirical estimates of a key parameter governing travel behavior, and to leverage these estimates to provide a glimpse of the possible energy impacts of vehicle connectivity and automation.

## 2. A Model of Private Vehicle Driving Decisions

Conceptually, vehicle ownership and driving decisions are a function of many factors: vehicle capital cost, the marginal cost of VMT (including fuel, time, and depreciation), and fixed costs of insurance and maintenance – collectively referred to as the total cost of ownership (TCO) Conceptually, vehicle ownership and driving decisions are a function of many factors: vehicle capital cost, the running costs of VMT (including fuel, time, maintenance, and depreciation), and fixed costs of insurance, registration fees and tolls – collectively referred to as the total cost of ownership (TCO) [26], the perceived cost of in-vehicle time, the utility an individual derives from travel, which depends on the goods and services obtained through travel, vehicle attributes, and individual preferences; and constraints such as income and time. In keeping with an extensive literature on empirical rebound effects (see, for example [14,18,27]), we focus our analysis specifically on the *marginal cost* of VMT conditional on vehicle choice. Marginal fuel and time costs are economically important and technologically relevant: together, they make up the majority of the variable cost of travel (19% and 45%, respectively [28]), and they are both projected to drop significantly with the diffusion of CAV technology [2,26,29,30]. Moreover, available data on these fuel and time costs (as well as VMT itself) allow us to develop empirically-grounded forecasts of CAVs' potential impact on energy use even when CAVs themselves have not yet been deployed commercially.

We begin by modeling VMT as a choice made by a utility-maximizing household, given constraints on income and time. Similar models exist in the energy rebound effect literature, but these do not include a time constraint [14,16,31], because energy efficiency improvements alone do not generally affect the use of time spent in a vehicle. In contrast, vehicle automation will decrease the opportunity cost of time through reduced in-vehicle attention requirements, which has the potential to alter driving decisions considerably. To capture this change, we adapt Linn's (2013) model of VMT choice [17] by adding a second constraint on time, following seminal economic theory on the allocation of time by Becker (1965) [32].

Consider a household that derives utility (U) from vehicle miles traveled ($VMT$) and consumption of a numeraire good ($y$), which proxies for all other goods in the economy. The household chooses levels of these variables subject to its available income and time as well as the monetary and time costs of $VMT$ and $y$. We write the maximization problem as follows:

$$\underset{VMT,y}{\text{MAX}}\ U(VMT, y) \tag{1}$$

such that:

$$P_f VMT + y \leq W \tag{2}$$

$$T_{vmt} + T_y + T_w \leq T \tag{3}$$

In Equation (2), $P_f$ is the per-mile fuel cost of $VMT$, while the price of $y$ is normalized to one; $W$ is household income. In Equation (3), $T_{vmt}$ is total travel time, $T_y$ is the consumption time of good $y$, $T_w$ is time spent on wage work, and $T$ is total available time. Total income $W$ is the product of $T_w$ and earned wage ($\tilde{w}$): $W = T_w \tilde{w}$. Similarly, $T_{vmt} = t_{vmt} VMT$ and $T_y = t_y y$, where $t_{vmt}$ and $t_y$ are the time input required per unit consumption of the two goods.

In equilibrium, the two budget constraints will be binding. We rewrite Equation (3) as

$$T_w = T - t_{vmt}VMT - t_y y \qquad (4)$$

and substitute this expression into Equation (2) to yield a single budget constraint:

$$(P_f + t_{vmt}\tilde{w})VMT + (1 + t_y\tilde{w})y = T\tilde{w} \qquad (5)$$

This single constraint follows from the fact that time can be converted to money through wage work. In other words, the opportunity cost of time spent on consumption is the income one forgoes in order to consume. Equation (5) expresses time in dollars: $t_{vmt}\tilde{w}$ is the dollar value of time spent on $VMT$, $t_y\tilde{w}$ is the analogous value for $y$, and $T\tilde{w}$ is the income one would have if all available time was devoted to work. The household spends its total "achievable" income either directly through expenditure on goods or indirectly by using time at consumption instead of work.

To derive an estimable equation for VMT choice, we must specify an explicit utility function. The household's true utility function is unknowable; we thus follow Linn (2013) [17] – whose goal is to estimate the energy rebound effect for passenger vehicles – and define utility as follows:

$$U(VMT, y) = -(VMT \cdot \xi)^\alpha + y \qquad (6)$$

where $\alpha < 0$ is a utility parameter and $\xi$ is vehicle quality which is known to the household but unobserved by the econometrician. Utility therefore increases in $VMT$ and vehicle quality. The chosen functional form is part of a class of utility functions that produce a constant price elasticity of demand, as we show below. While constant demand response is a special case and unlikely to hold in reality, it is nonetheless useful here to clearly demonstrate how fuel and time costs affect VMT demand.

The optimum choice of $VMT$ and $y$ satisfies the first-order condition:

$$\frac{\partial U}{\partial VMT} = -\alpha\xi(VMT \cdot \xi)^{\alpha-1} + \frac{\partial y}{\partial VMT} = 0 \qquad (7)$$

Using the budget constraint (Equation (5)), we can express $y$ as a function of $VMT$ and parameters. Substituting this expression into Equation (7), rearranging terms, and taking the logarithm of both sides yield:

$$\log(VMT) = \left[ \frac{1}{1-\alpha} \log(-\alpha) + \frac{\alpha}{1-\alpha} \log(\xi) + \frac{1}{1-\alpha} \log\left(1 + t_y \widetilde{w}\right) \right]$$
$$- \frac{1}{1-\alpha} \log(\pi_{vmt}) \tag{8}$$

where we define $\pi_{vmt} = P_f + P_t = P_f + t_{vmt}\widetilde{w}$ as the time-inclusive marginal cost (or price) of travel. Since $\alpha < 0$ , Equation (8) implies that $VMT$ decreases with higher $\pi_{vmt}$. The log-log form of this equation makes the coefficient on $\pi_{vmt}$, $(\frac{-1}{1-\alpha})$, interpretable as a first-order approximation of the elasticity of $VMT$ with respect to $\pi_{vmt}$. Denoting this elasticity by $\varepsilon_{vmt}$ and collecting the first three terms of Equation (8), we have:

$$\log(VMT) = \varepsilon_{vmt} \log(\pi_{vmt}) + constant \tag{9}$$

With data on VMT, fuel economy, gasoline prices, and travel time cost, we can fit this equation and estimate the key parameter of interest, $\varepsilon_{vmt}$.

## 3. Data and Empirical Strategy

### 3.1. Data

We obtain data on the price and quantity of VMT from the National Household Travel Survey (NHTS) [25]. This representative nationwide survey is conducted by the Federal Highway Administration (FHWA) in order to assist policymakers and transportation planners in understanding travel behavior and how it changes over time. Our main source is the 2017 round of the NHTS, but we test the robustness of our results to use of the 2009 round as well. In both of these surveys, households submit day-long travel logs which include VMT and time spent driving for each vehicle driven. FHWA then imputes annual totals from these daily numbers using weight adjustments. Respondents also report the make and model of each vehicle, as well as the price of retail gasoline on the day of reporting. In addition to providing these vehicle data, the NHTS records several socioeconomic and demographic characteristics of households. The full sample includes 129,696 observations; our analysis sample consists of the 114,923 households with non-missing values for our key analysis variables.[5] In all analyses, we use sampling weights provided in the NHTS and equal to the reciprocal of selection probability to make the sample nationally representative.[6]

Table 1 summarizes the household-level NHTS variables on which we draw to construct our analysis. We tabulate means and standard deviations, both overall and within each of five specific income groups. While before-tax household income is reported in eleven distinct intervals in the 2017 NHTS, we follow Wadud (2017) [26] and collapse intervals into five income groups with roughly the same number of households. Sample-average annual VMT is 16,254 miles and rises monotonically from the first (i.e., lowest) income group to the fifth (highest); the latter group drives more than 2.5 times as many miles as the former. Annual driving time follows a similar pattern but drops slightly from the fourth income group to the fifth. Reported gas prices rise monotonically in income group but only differ by about five cents per gallon from the first income group to the fifth. Average fuel economy, weighted by miles traveled in each one of a household's vehicles, exhibits an inverse U-shaped relationship with income group.

---

[5] We remove the 3.1% of households with unreported income and an additional 8.4% who report zero VMT, no vehicle ownership, a vehicle model from before 1984 (which is not included in the EPA testing data), or unknown vehicle make and model.

[6] Analysis without weights would yield internally valid estimates of our parameters of interest but would not be nationally representative.

Table 1. Summary statistics for 2017 NHTS (non-exhaustive list of variables)

| Variable | U.S. Average | 1st Income Group | 2nd Income Group | 3rd Income Group | 4th Income Group | 5th Income Group |
|---|---|---|---|---|---|---|
| Income Interval | - | Up to $24,999 | $25,000 to $49,999 | $50,000 to $74,999 | $75,000 to $124,999 | Over $125,000 |
| Average Income[†] | $70,237 | $19,447 | $40,976 | $64,563 | $106,173 | $180,674 |
| Annual VMT (Miles) | 16,254 (20,166) | 8,592 (14,447) | 14,146 (17,818) | 17,580 (20,528) | 20,589 (21,879) | 22,055 (22,870) |
| Annual Driving Time (Hours) | 482.18 (496.11) | 269.73 (302.21) | 434.27 (455.73) | 521.69 (537.89) | 615.38 (622.23) | 601.67 (598.75) |
| Reported Gas Price ($/gallon) | 2.392 (0.2066) | 2.3747 (0.2018) | 2.384 (0.2026) | 2.3902 (0.2061) | 2.4013 (0.2076) | 2.4225 (0.212) |
| Weighted Average Fuel Economy (MPG)[∇] | 23.69 (10.99) | 23.11 (10.41) | 24.90 (12.21) | 25.30 (11.10) | 24.41 (10.95) | 23.16 (13.11) |
| Household Size (Persons) | 2.514 (1.380) | 2.146 (1.451) | 2.273 (1.325) | 2.532 (1.363) | 2.776 (1.324) | 2.987 (1.233) |
| Count of Adults | 1.925 (0.821) | 1.623 (0.843) | 1.804 (0.807) | 1.959 (0.799) | 2.101 (0.767) | 2.215 (0.733) |
| Count of Drivers | 1.762 (0.882) | 1.205 (0.852) | 1.623 (0.790) | 1.842 (0.804) | 2.049 (0.796) | 2.210 (0.783) |
| Count of Vehicles | 1.935 (1.255) | 1.130 (0.970) | 1.727 (1.067) | 2.078 (1.169) | 2.357 (1.237) | 2.545 (1.306) |
| Indicator for urban area (1 = urban; 0 = rural) | 0.808 (0.378) | 0.834 (0.363) | 0.817 (0.385) | 0.801 (0.394) | 0.818 (0.385) | 0.857 (0.348) |
| Census Tract Population Density (Persons per square mile) | 5,647 (7,345) | 6,314 (7,816) | 5,388 (6,897) | 5,340 (7,180) | 5,273 (7,084) | 6,005 (7,772) |
| Census Tract Housing Density (House per square mile) | 3,042 (5,465) | 3,386 (5,529) | 2,850 (4,978) | 2,809 (5,115) | 2,812 (5,369) | 3,452 (6,461) |
| N | 114,923 | 22,959 | 25,793 | 21,45 | 26,005 | 19,531 |

Standard deviations are reported in parentheses. All observations are weighted using the sample weights provided in the NHTS.

[†] Average income within income group is calculated from the 2016 Consumer Expenditure Survey.

[∇] Fuel economy is derived from EPA Fuel Economy Testing Data [33] for vehicles.

To produce a fuel price of VMT ($P_f$ in dollars per mile) for each household, we multiply its reported fuel price per gallon by its weighted average fuel economy:

$$P_f = \frac{\phi}{\sum_{i=1}^{n} VMT_j} \sum_{j=1}^{n} \frac{VMT_j}{MPG_j} \tag{10}$$

where $n$ is the number of vehicles that a household uses, $VMT_j$ and $MPG_j$ are vehicle miles traveled and fuel economy (miles per gallon) of the $j$th vehicle, respectively, and $\phi$ is the price of gasoline (dollars per gallon). Unlike the 2009 NHTS, the 2017 NHTS does not itself report vehicle fuel economy; we thus obtain combined MPG (45% city, 55% highway) from EPA Fuel Economy Testing Data [33] for all vehicles in our sample.[7]

The time component of the marginal cost of travel ($P_t$), which we refer to as travel time cost (TTC), is not directly observable in NHTS data, nor in any other dataset of which we are aware. To overcome this data problem, we follow the economics literature and the U.S. Department of Transportation's (US DOT) 2016 guidelines for Revised Value of Travel Time [34] and parameterize TTC as a function of wage. The NHTS only reports an annual income bracket for each household; we calculate the "equivalent" hourly wage of each household by dividing the average income in a household's bracket, taken from the 2016 Consumer Expenditure Survey, by 2,080 working hours in a year. Like Chen et al. (2016), we then categorize all survey-reported trips as either "work-related" or "non-work", the latter of which includes shopping, family/personal errands, school/church visits, social/recreational trips, among others [35]. We value work-related trips at 100% of hourly wage and non-work trips at 50% of hourly wage, following US DOT guidelines [34].[8] Finally, we compute a weighted average of these trip values using time shares of each trip type as weights:

$$P_t = \frac{\left(\gamma_W \, \hat{w} + \frac{1}{2}\gamma_{NW}\hat{w}\right) \times \sum T_{vmt}}{\sum VMT} \tag{11}$$

Here, $\gamma_W$ is the share of total travel time devoted to work-related trips, $\gamma_{NW}$ is the corresponding share for non-work trips, $\hat{w}$ is imputed hourly wage, and $\sum T_{vmt}$ is the total time spent on all trips. While our focus is on the travel time cost per mile, we also plot the time cost per hour in the Appendix (Figure A1). In our sample, the average time cost per hour of travel is 19.56 $/h, which is comparable to the Value of Travel Time recommended by US DOT (18 $/h) [34].

---

[7] Although, the EPA fuel efficiency data is known to overstate fuel economy of vehicles, it is the most comprehensive dataset available.

[8] In the appendix, we show results of a robustness check in which we use alternative definitions of travel time cost.

Figure 1 displays fuel, time, and aggregate marginal costs by income group. The aggregate marginal cost of VMT ($\pi_{vmt}$) rises steeply and monotonically with income group, as does the time cost component ($P_t$). The fuel component ($P_f$) shows a shallow U-shaped relationship with income group. The time cost generally dominates the fuel cost, consistent with previous research that highlights the relative importance of travel time cost [7,26,29]. In our sample, both time cost and aggregate cost per mile rise faster than linearly in income group.[9] In fact, the top income group has nearly seven times the travel time cost as the bottom income group and more than three times the aggregate marginal cost of travel.

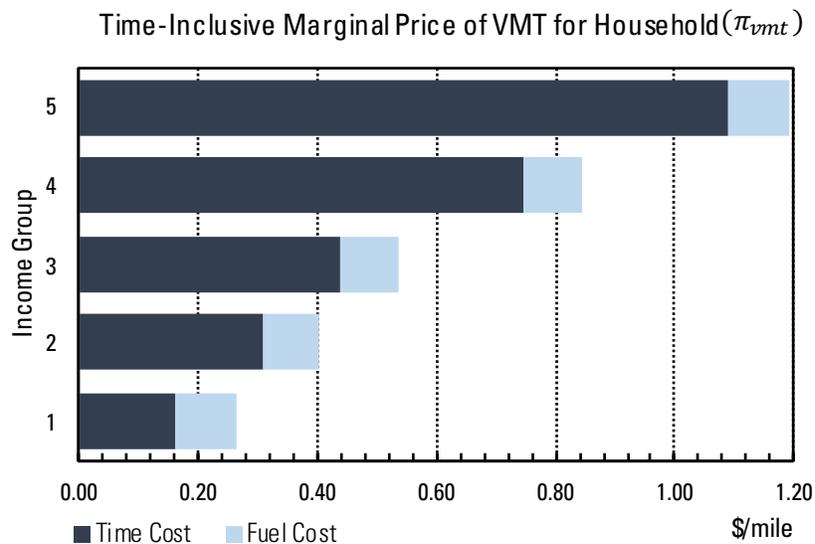

Figure 1. Marginal price of one vehicle mile traveled (VMT) by income group for the average household in each income group. Equations 10 and 11 are used to derive fuel cost and time cost per mile of driving.

---

[9] This is a result of defining time costs as proportional to income, as well as the non-linear relationship between median income and our chosen income grouping.

## 3.2. Empirical Estimation

Using the above data, we fit various specifications of Equation (9) to estimate the price elasticity of demand for VMT. We choose four closely-related econometric models:

**Model 1:**     $\log(VMT_i) = \beta_0 + \beta_1 \log(P_{f,i}) + \gamma \vec{X}_i + \omega_i$     (12)

**Model 2:**     $\log(VMT_i) = \beta_0 + \beta_2 \log(P_{t,i}) + \gamma \vec{X}_i + \omega_i$     (13)

**Model 3:**     $\log(VMT_i) = \beta_0 + \beta_1 \log(P_{f,i}) + \beta_2 \log(P_{t,i}) + \gamma \vec{X}_i + \omega_i$     (14)

**Model 4:**     $\log(VMT_i) = \beta_0 + \beta_3 \log(\pi_{vmt,i}) + \gamma \vec{X}_i + \omega_i$     (15)

The subscript $i$ indexes a household. $VMT_i$, $P_{f,i}$, $P_{t,i}$, and $\pi_{vmt,i}$ are as described in Section 2. $\vec{X}_i$ is a vector of household characteristics taken directly from the NHTS. One subset of this vector pertains to household members and includes household size, number of adults and drivers, indicators for respondent's race, and indicators for a household's age distribution.[10] A second subset contains socioeconomic measures including indicators for income group and homeownership as well as a count of a household's vehicles. A third pertains to location and includes census block group population density and housing density, indicators for urban (versus rural) area and metropolitan statistical area (MSA), MSA size, and indicators for values of a categorical variable defined by census division, whether or not a MSA has a population above one million, and whether or not an MSA has a subway system. A fourth, and final, subset includes indicators for survey month of year and day of week. We choose these control variables to match Linn (2013) and Su (2012) [17,18] as closely as possible. Lastly, $\omega_i$ is an error term that captures the effect of unobserved drivers of VMT.

We estimate each model via Generalized Least Squares regression, using the sampling weights provided by the NHTS. We cluster standard errors by MSA, to allow for correlation of individual errors within each MSA. The log-log functional form has three virtues: it is motivated directly by our model in Section 2; it gives the coefficient on $\log(\pi_{vmt,i})$ the interpretation of the price elasticity of demand for VMT; and, in our specific empirical context, it produces model residuals that are normally distributed, implying that heteroscedasticity is of minimal concern.

---

[10] Indicators for a household's age distribution include, for instance, "two or more adults, youngest child 16-21".

Model 1 specifies VMT to be a function of only the fuel component of VMT price (i.e., not the corresponding time component). This specification is typical in the economics literature on energy efficiency rebound and yields an estimate of VMT elasticity with respect to the fuel price of VMT ($\hat{\beta}_1 = \hat{\varepsilon}_f$). However, it is susceptible to omitted variable bias if the omitted time component of price is correlated with the included fuel component. Model 2 is the time-cost analog of Model 1; it yields a VMT elasticity with respect to the time cost of VMT ($\hat{\beta}_2 = \hat{\varepsilon}_t$) and suffers from the same risk of omitted variable bias. Models 3 and 4 mitigate this risk by including the costs of both fuel and time as explanatory variables. Model 3 allows for joint estimation of the fuel-price and time-cost elasticities, $\hat{\varepsilon}_f$ and $\hat{\varepsilon}_t$. Parameter estimates from this model can be compared to those of Models 1 and 2 to quantify the bias of the latter.

Model 4 is the specification of VMT that follows directly and exactly from our economic model of VMT choice in Section 2. Fitting this model yields an estimate of the average combined, fuel- and time-inclusive price elasticity of VMT, $\hat{\varepsilon}_{vmt}$. This combined elasticity is related to $\hat{\varepsilon}_f$ and $\hat{\varepsilon}_t$ but not necessarily a linear function of the two. If $\hat{\varepsilon}_f \neq \hat{\varepsilon}_t$, then $\hat{\varepsilon}_{vmt}$ will depend intrinsically on the relative magnitudes of changes in $P_f$ and $P_t$. In the special case in which $P_f$ and $P_t$ change by the same proportion, $\hat{\varepsilon}_{vmt} = \hat{\varepsilon}_f + \hat{\varepsilon}_t$; but in the general case where cost changes are not equal in proportion, $\hat{\varepsilon}_{vmt}$ may be larger or smaller than the sum of $\hat{\varepsilon}_f$ and $\hat{\varepsilon}_t$.

Income plays an especially important role in the determination of travel behavior and therefore transportation equity. As our theoretical model shows, VMT demand is affected by income through both the income budget constraint (i.e., money available to pay for VMT) and the time budget constraint (i.e., the opportunity cost of time, which depends on wage). As such, we break out our estimation of Models 1-4 by income group, interacting our price variables with indicators for income group[11]. In all cases, we omit the interaction of price with the lowest income-group indicator, so that the point estimate on the (uninteracted) price level is interpretable as the elasticity corresponding to this bottom group.

## 3.3. Scope and Limitations

Our theoretical model and empirical strategy are well-suited to leverage household-level driving data to estimate demand elasticities, but they abstract from several qualitatively important aspects of driving decisions. First, we do not model the capital decision of vehicle purchase. A static, two-period economic model with a first stage capturing vehicle purchase would show that buying a new car tightens the budget constraint and thus pushes VMT downwards [31,36]. This, in turn, would suggest that our elasticity estimates will be biased upwards. In a dynamic model, on the other hand, a forward-looking consumer might not adjust VMT in response to the (planned and expected) expense of a new car. More generally, the upfront cost of CAV use will depend on future innovation in CAV production technology as well as the prevalence of shared CAV modes. In any case, since we estimate elasticities by comparing changes in marginal costs, the external validity of these estimates rises as the upfront cost of CAV use decreases.

We also note that our measurement of costs includes fuel and time but not depreciation, maintenance, insurance, or congestion. Our omission of depreciation, maintenance, and insurance costs is motivated by a lack of data on these cost components and little consensus on the changes likely to occur with CAV technology diffusion along these dimensions. We note, however, that bias from omission of these variables is only a risk insofar as changes in depreciation and insurance costs are correlated with changes in fuel and time costs. Congestion is similarly unobservable in our data and difficult to forecast in a CAV-dominant mobility paradigm. Every additional VMT comes with an external congestion cost to other drivers that we do not measure. At low levels of CAV penetration, congestion costs may be negligible, but at higher levels, and with large associated reductions in the marginal cost of travel, congestion may be an important check on induced travel [37].

---

[11] Our primary objective in this paper is to estimate average elasticities, both overall and within income group. For applications that benefit from more disaggregated predictions, machine learning and artificial intelligence methods may provide significant gains in precision. For instance, these methods are increasingly being used to predict household-level electricity demand as a function of observable characteristics [50–52].

Finally, our travel time cost measure is imputed from reported income data. It is thus subject to significant measurement error as well as a risk of omitted variable bias. We see our imputation, which follows a long literature in economics and transportation research that links opportunity costs to wage, as the best we can do to estimate the opportunity cost of time spent traveling. Measurement error biases estimates towards zero; on the other hand, if households that drive more also value time more for reasons other than income, the omission of such explanatory factors might bias our estimates away from zero. It is for this latter reason that we include a large vector of control variables in regression. Ultimately, we make no strong claim on the statistical precision of our estimates; rather, we argue that our exercise illustrates the sizeable role that time cost plays in current travel decisions and will play in a future with driverless vehicles.

## 4.  Estimates of Price Elasticity of Demand for VMT

Table 2 displays our estimates of the sample-wide elasticity of demand for VMT with respect to different components of VMT price. The point estimate obtained from Model 1 implies a fuel price elasticity of approximately -0.14; that is, a one percent rise (drop) in the fuel price per VMT is associated with a 0.14 percent drop (rise) in VMT itself. This magnitude is well within the range provided in the existing literature [14,17–20] , which includes estimates as low as -0.06 [18,19]  and as high as -0.28 [20]. Model 2, meanwhile, yields a corresponding point estimate of approximately -0.45 for the time cost elasticity. While this is significantly larger than our fuel price elasticity estimate, such a large difference is consistent with the findings of the travel demand literature [10,11,13,38]. There are few existing estimates of the elasticity of VMT with respect to travel time cost, and there is no consensus on its magnitude.

Our estimates from Models 1 and 2 are susceptible to omitted variable bias, because each omits one of the two key components of the marginal cost of travel. In fact, $P_f$ and $P_t$ are positively correlated in our data (the Pearson correlation coefficient is 0.37), which implies that our estimates from Models 1 and 2 are biased upwards. Our results from Model 3 confirm this: the jointly estimated fuel and time price elasticities are approximately -0.10 and -0.40, respectively, and both are smaller than their separately-estimated analogs.[12] Together, our results using Models 1-3 suggest that existing estimates of travel demand elasticities may be systematically biased upwards. We know of no studies that jointly consider fuel prices and the opportunity cost of time in empirical measurement of elasticities. This is primarily due to a lack of available data on the value of time [7], which is a challenge for us just as much as any other researchers. While we do not know households' true valuations of time, there is broad consensus that the opportunity cost of travel rises with income [7]. As long as the fuel price of VMT rises in income, as it does in our case, omitting one cost component or the other will produce upward bias in elasticity estimates.

---

[12] A neoclassical economic model would yield the prediction that $\hat{\varepsilon}_f = \hat{\varepsilon}_t$. The fact that this is not the case in our context suggests the possibility that some behavioral-economic phenomenon causes households to respond differently to a change in fuel cost than a dollar-equivalent change in time cost.

Table 2. Results of elasticity estimation (main explanatory variables) for different models

|  | Model 1 | Model 2 | Model 3 | Model 4 |
|---|---|---|---|---|
| $\hat{\varepsilon}_f$ | -0.1408*** (0.028) | - | -0.0989*** (0.017) | - |
| $\hat{\varepsilon}_t$ | - | -0.4486*** (0.042) | -0.4007*** (0.048) | - |
| $\hat{\varepsilon}_{vmt}$ | - | - | - | -0.3920*** (0.049) |
| Pseudo $R^2$ | 0.227 | 0.261 | 0.272 | 0.240 |

The dependent variable is $\log(VMT)$. Each column reports a separate regression. All regressions include fixed effects and control variables described in Section 3.2. Observations are weighted by the household sample weights. Asterisks denote 1 (***), 5 (**), and 10 (*) percent significance levels.

Model 4, like Model 3, accounts for both the fuel price and the time price; however, it parameterizes demand to depend only on the (log) sum of the two, rather than each individually. Using this model, we estimate a combined elasticity of demand ($\hat{\varepsilon}_{vmt}$) of approximately -0.39. Since $\hat{\varepsilon}_f$ and $\hat{\varepsilon}_t$ from Model 3 are markedly different, there is no special reason to believe that $\hat{\varepsilon}_{vmt}$ is equal to the sum of $\hat{\varepsilon}_f$ and $\hat{\varepsilon}_t$. Rather, the relationship between these three parameters depends on the empirical distribution of prices in our particular context. In this case, the time channel dominates the fuel channel, as $\hat{\varepsilon}_{vmt}$ is approximately the same as $\hat{\varepsilon}_t$. To us, this comparison exercise underscores the importance of using separate fuel and time price elasticities in travel demand forecasts. Our combined price elasticity estimate is internally valid, but it is unlikely to be externally valid to scenarios in which the relative prices and price changes pertaining to fuel and time are different.

Our estimated combined VMT elasticity of -0.39 differs significantly from other estimates in the existing literature. This discrepancy illustrates the importance of empirical analysis in the calibration of demand response. Elasticities of travel demand are a key input into any forecast of CAV travel and energy use; one must be careful in applying estimates from one context to another, different context. Using existing fuel price elasticity estimates – which are 25-85% lower than our combined elasticity [14,17–20] – to predict energy rebound would almost certainly underestimate the impact of vehicle automation on energy use. On the other hand, using previously published estimates of VMT elasticity with respect to generalized travel costs – which are 60-400% higher [4,5] than ours – would very likely *overestimate* the energy use impact of CAVs.

It is not just the type of price change (fuel- or time-specific) that dictates the size of the demand response; it is also household wealth that matters. Table 3 displays the results of estimating modified versions of Models 3 and 4 that allow for differences in demand response across the wealth spectrum. Panel A contains our individual fuel and time price elasticities, while Panel B contains our combined price elasticities. Figure 2 shows the same results graphically. There is significant heterogeneity in all three parameter estimates across income groups.

Panel A of Table 3, which reports results from Model 3, show that the gap between $\hat{\varepsilon}_f$ and $\hat{\varepsilon}_t$ in the overall sample persists within each income group as well. Panel B of Table 3, which reports results from Model 4, reveals the relationships between wealth and demand response to specific components of VMT price. The absolute-value fuel price elasticity drops in wealth until the last income group; in contrast, the absolute-value time cost elasticity *rises* monotonically in wealth. These findings imply that richer households have less elastic demand than poorer ones with respect to fuel price changes and more elastic demand with respect to time cost changes. We do not attempt to explain these findings here, but we note that both positive and negative relationships between demand elasticity and wealth have been found in the existing economics literature [19,39–41]. On the one hand, wealthier households may engage in more discretionary travel than poorer ones, and for that reason their demand for VMT may be more elastic to price. On the other hand, wealthier households are also generally less price-sensitive than poorer ones, and this may make their demand less elastic. Our results using Model 4 (Table 3, Panel B) reveal that, on aggregate, wealthier households in our context have relatively more elastic demand for VMT. For all four models, the signs and relative magnitudes of estimated coefficients on control variables are consistent with both economic intuition and the findings of previous studies utilizing similar approaches and datasets [17,18].

Table 3. Elasticity estimates by income group

| Income Group | 1st Income Group | 2nd Income Group | 3rd Income Group | 4th Income Group | 5th Income Group |
|---|---|---|---|---|---|
| **Panel A: Model 3** | | | | | |
| $\hat{\varepsilon}_f$ | -0.153*** | -0.131*** | -0.097*** | -0.092*** | -0.109*** |
| | (0.026) | (0.012) | (0.019) | (0.015) | (0.017) |
| $\hat{\varepsilon}_t$ | -0.290*** | -0.403*** | -0.446*** | -0.463*** | -0.474*** |
| | (0.063) | (0.055) | (0.049) | (0.038) | (0.048) |
| **Panel B: Model 4** | | | | | |
| $\hat{\varepsilon}_{vmt}$ | -0.256*** | -0.351*** | -0.401*** | -0.444*** | -0.421*** |
| | (0.048) | (0.052) | (0.051) | (0.037) | (0.042) |

The dependent variable is $\log(VMT)$. Both regressions include fixed effects and control variables described in Section 3.2. Observations are weighted by the household sample weights. Asterisks denote 1 (***), 5 (**), and 10 (*) percent significance levels. The pseudo $R^2$ of regression for Panel A is 0.272 and for Panel B is 0.240.

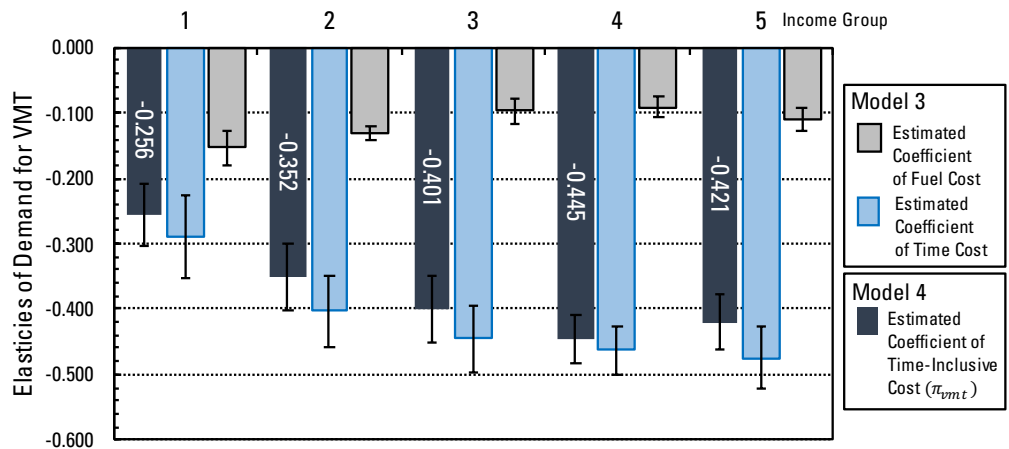

Figure 2. Estimated elasticities of demand with respect to $P_t$ and $P_f$ (from Model 3) and $\pi_{vmt}$ (from Model 4). Clustered standard errors are shown as error bars. Standard errors are clustered by MSA, and observations are weighted by the household sample weights.

We conduct two sets of robustness checks to assess the sensitivity of our results to key modeling decisions. First, we compare results of using the 2017 NHTS to those of using the 2009 NHTS while maintaining the same definitions and parameterizations wherever possible.[13] Appendix Table A1 displays our findings, including sample-wide and income-group specific estimates. The absolute magnitudes of all three sample-wide elasticity estimates are modestly larger in 2009 than in 2017, as highlighted in Column 7. Across income groups, trends in $\hat{\varepsilon}_t$ and $\hat{\varepsilon}_{vmt}$ are consistent in both the 2009 data and the 2017 data, while $\hat{\varepsilon}_f$ exhibits more of a U-shaped relationship with income in the 2009 data. Some variation in estimates across the two survey rounds is expected, since baseline income, fuel prices, and fuel economy are not constant over time. In fact, 2009 is notably defined by the onset of the Great Recession. The fact that 2009 elasticity estimates are qualitatively similar to our main, 2017-based estimates lends credence to our empirical strategy and results.

In the second robustness check, we test how the definition of time cost affects estimation results. We employ two alternative definitions of travel time cost: first, that it is equal to 100% of hourly wage for all trips; and second, that it is equal to 50% of hourly wage for all trips (Appendix Figure A2). We report the results in Appendix Table A2. Mechanically, the first of these definitions causes estimated time and combined price elasticities to fall relative to our preferred estimates, while the second causes estimated elasticities to rise. The former effect is much more pronounced than the latter, perhaps because the high proportion of non-work trips in our data makes our preferred estimates much more similar to alternative definition 2. Meanwhile, trends in all three elasticity parameter estimates (not shown for fuel prices) across income groups are robust. While our alternative definitions rely on reported income just as much as our preferred estimate, this robustness check does imply that our qualitative findings are not solely an artifact of defining work and non-work trips differently.[14]

---

[13] Household income groupings in the raw 2009 NHTS do not exactly match those in the 2017 NHTS. We aggregate income groups in the 2009 data to match those of the 2017 data as closely as possible.

[14] We additionally conduct several robustness checks to assess the sensitivity of results to model specification and parametrization. All results are within a reasonable range of our main estimates.

# 5.  Forecasting CAV-Induced Travel and Energy Use

One way to predict the travel and energy impacts of CAVs is by estimating the demand response to changes in energy efficiency and travel time cost that may occur as a result of CAV technology. The two primary inputs to such an analysis are travel demand elasticities and price changes. We use our estimates from Section 4 for the former and a range of estimates based on the existing CAV literature for the latter. While it is widely understood that automation and connectivity will enable a range of fuel-saving practices at the vehicle level, estimates of the magnitude of associated fuel and time cost changes are rare and largely speculative. Studies collectively suggest 5% to 20% energy efficiency improvement in CAVs compared to conventional counterparts, mainly due to optimal driving cycle, eco-routing, congestion reduction, and improving vehicle electrification[15] attributes [2–5,24,37].

Reductions in TTC for CAVs relative to conventional cars are predicted to come mainly from decreased attention demands and driving-related stresses [5], the resulting increase in opportunities to engage in alternative in-vehicle activities[16] [42,43], and increases in travel speeds (through improved safety and traffic flow) [44]. Comparing previous studies of TTC in rail travel versus vehicle travel, Wadud (2017) estimates that the switch from conventional to CAVs will yield a 25-60% reduction in TTC [26]. The recent survey results of Correia et al. (2019) show that a CAV with an office interior could reduce travel time cost by 26% compared to a conventional car [45]. 60% is consistently accepted as the upper bound of possible TTC reductions in the literature [4,5,22,30,42,44], since in-vehicle attention requirements cannot be completely eliminated.[17]

---

[15] While the effect of vehicle electrification on net energy consumption is similar to fuel economy improvement, it could have a much different impact on vehicle tailpipe emissions as well as upstream emissions from electricity generation.

[16] Such activities include, for example, watching movies, sleeping, eating, working, checking emails, browsing web and social media.

[17] Some studies argue that increased productivity while riding with CAVs is not guaranteed. Apprehension [53] or motion sickness may limit the ability of passengers to engage in other activities or raise the disutility of travel [43,54]. Short average trip times may not provide sufficient time for sustained productivity or sleep [53].

In our forecasting exercise, we increase fuel economy ($MPG$) and travel time cost ($p_t$) by $X$ and $Y$, respectively, where $X \in [0.05, 0.2]$ (or 5-20%) and $Y \in [0, 0.6]$ (or 0-60%). The direct outcome of interest is the travel demand induced by CAV cost changes as a percentage of the pre-CAV "business as usual (BAU)" ($\delta = \frac{VMT_{CAV}}{VMT_{BAU}} - 1$). We use our fitted regression function from Model 3 to generate VMT predictions for any cost conditions: $\widehat{VMT} = e^{\hat{\beta}_0} p_f^{\hat{\varepsilon}_f} p_t^{\hat{\varepsilon}_t}$. Substituting our expression for $\widehat{VMT}$ into our equation for $\delta$, rewriting $p_f = \phi / MPG$, and assuming gasoline price $\phi$ is fixed, we obtain:

$$\delta = \left(\frac{MPG_{BAU}}{MPG_{CAV}}\right)^{\hat{\varepsilon}_f} \left(\frac{p_{t_{CAV}}}{p_{t_{BAU}}}\right)^{\hat{\varepsilon}_t} - 1 \qquad (16)$$

Finally, we re-express CAV values as functions of BAU using $X$ and $Y$ and simplify to yield

$$\delta = \left(\frac{1}{1+X}\right)^{\hat{\varepsilon}_f} (Y)^{\hat{\varepsilon}_t} - 1 \qquad (17)$$

We compute $\delta$ overall (using elasticities from Column 3 in Table 2) and for each income group (using elasticities from Columns 1-5 in Table 3), iterating over values of $X$ and $Y$ in increments of 0.05.

In principle, we could use elasticity estimates from any of our four empirical models (Equations 12-15) to forecast induced travel. We prefer to use Model 3 estimates because they strongly suggest that demand response depends on the specific source of price changes (fuel vs. time). Models 1 and 2 consider only one source or the other and are thus relatively more susceptible to omitted variable bias. Model 4 accounts for both fuel cost and time cost, but it does not allow the elasticity of demand to vary with the relative sizes of fuel and time cost changes.[18] Consider any two different $\{X, Y\}$ pairs that, on aggregate, produce the same proportional change in $p_t$: Model 3's results strongly suggest that these two pairs produce different VMT demand response; using Model 4 would force them to yield the same response.

---

[18] The Model-4 equivalent equation to Equation 16 is $\delta = \left(\frac{\pi_{vmt_{CAV}}}{\pi_{vmt_{BAU}}}\right)^{\hat{\varepsilon}_{vmt}} - 1$.

Motivated by this discrepancy, we show forecasting results based on Model 3 here and those based on Model 4 in the Appendix.

Figure 3 depicts our results in the form of heat maps. The x-axis indicates the fuel economy improvement, while the y-axis indicates the time cost reduction. Color depth measures the induced travel demand $\delta$ in percentage terms. Two patterns are readily observable. First, the magnitude of induced travel rises monotonically with increases in either $X$ or $Y$, consistent with negative price elasticities of demand. For the average household in the 2017 NHTS, our range of simulated price changes produces a minimum forecast of 2% induced travel and a maximum of 47%. Second, induced travel rises with income group for any given $(X, Y)$ pair, consistent with larger absolute-value time cost elasticities among richer households that dominate smaller absolute-value fuel price elasticities. In the lowest income group, the average household is forecast to increase VMT by 1-35%, while the corresponding range is 3-58% in the highest income group.

The dashed lines in Figure 3 connect forecasted induced travel to forecasted energy use. In particular, they indicate combinations of $(X, Y)$ that yield zero net change in energy use. Such an exact offsetting is possible because, even as fuel and time price drops induced travel, energy efficiency reduces the energy required per unit of travel. The slopes of the dashed lines therefore denote the rate at which time costs need to drop in order to fully offset the energy savings from an additional percentage rise in fuel economy. For instance, Figure 3 indicates that, in the sample-average household, a 20% rise in fuel economy would lead to net energy savings unless travel time cost drops by 38% or more. In each heat map, the area below and to the right of the dashed line is characterized by net decreases in energy use from the simulated changes, while the area above and to the left of the dashed line is characterized by net increases, i.e., what is known in the literature as "backfire" [31].

It is apparent, both overall and in each specific income group, that a wide range of CAV cost changes can produce backfire. Of course, not all combinations of $(X, Y)$ are equally likely to occur. We therefore do not argue that backfire is "likely" to occur at any specific levels of $X$ and $Y$. Our empirical analysis nevertheless suggests the possibility of net energy increases from changes which are well within the ranges predicted in the CAV literature. Furthermore, backfire is increasingly likely in higher income groups. This trend follows naturally from two empirical facts about relatively richer households in the 2017 NHTS: (1) a greater proportion of their imputed total travel costs come from time rather than fuel; and (2) they have more elastic demand with respect to time costs. We predict that the energy savings from a 20% rise in fuel economy can be offset by a 50% drop in travel time cost in the lowest income group; in the highest income group, however, only a 32% drop in time costs is needed.[19]

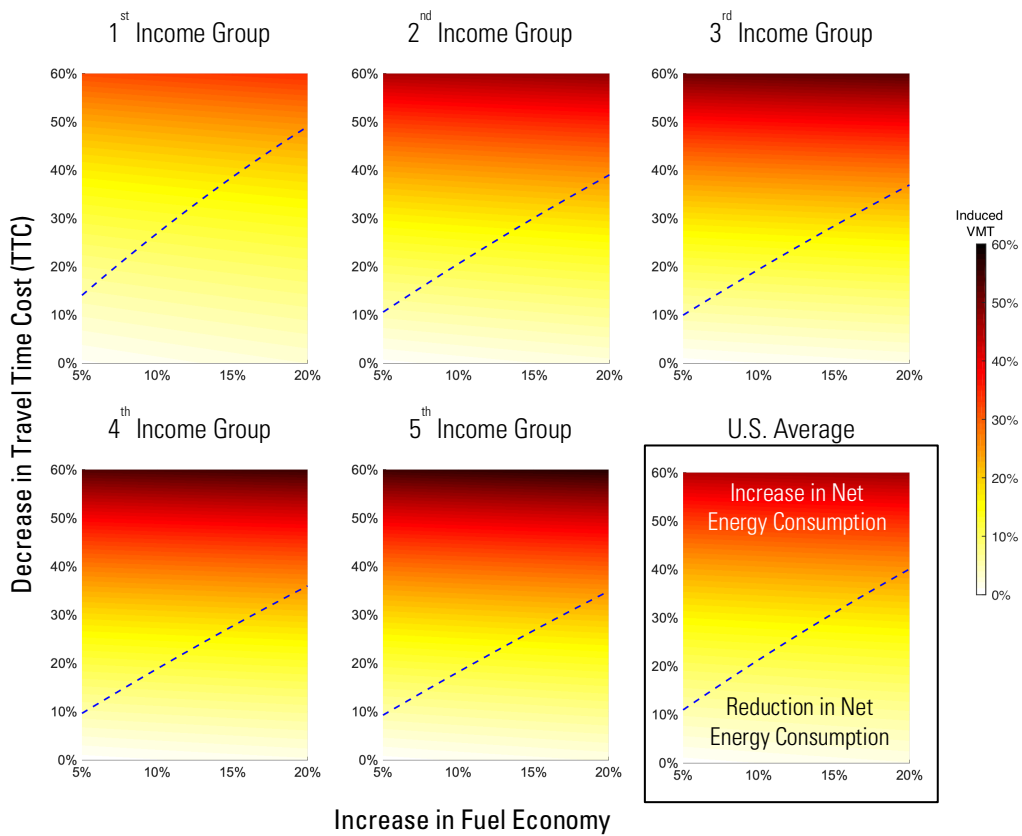

Figure 3. Simulation of induced travel to fuel economy improvement and reduction in TTC for CAVs, and the impact on net energy consumption. Any point above dashed curves represent the case of backfire (increase in net energy consumption despite increase in fuel economy).

---

[19] Appendix Figure A3 depicts our simulation results from use of Model 4. Overall induced travel demand is lower at any $\{X, Y\}$, and the slope of the dashed line changes more dramatically with income group. Otherwise, the patterns are the same.

There are other existing studies of the travel demand changes stemming from CAV technology. We highlight the methods and results of some of these in Table 4. In the prior literature, higher VMT in CAVs is attributed not just to higher passenger travel but also to, variously, new user groups [46], empty vehicle travel (i.e., unoccupied VMT) [47,48], and the possibility of shifts in mode choice and urban sprawl [4,35,49]. New user groups include minors and elderly and medically infirmed individuals who *may begin* traveling with the availability of CAVs. Empty vehicle travel refers to VMT with no passengers, such as what might occur in a private CAV before or after passenger drop-off or in a shared CAV dispatched to pick up the next passenger. Mode choice shift includes substitution of CAV use for public transit, and urban sprawl refers to the possibility of changes to residential location choice due to CAV availability. Our work focuses entirely on induced travel among existing drivers and yields estimates of overall VMT change in the range of 2 to 47 percent.

Table 4.  Literature estimate of changes in VMT due to CAV technology (list is non-exhaustive). For detailed discussion refer to [2].

| Study | Method | Estimate of VMT change | Sources of VMT change |
|---|---|---|---|
| Childress et al. (2015) [42] | Activity-based model for Puget Sound region | -30% to +20% | Changes in driving cost through value of travel time, road capacity, and parking cost |
| Fagnant and Kockelman (2015) [44] | Scenario-based analysis based on assumptions | +10 to +20% | Induced travel demand |
| Harper et al. (2016) [46] | Demand wedge analysis based on 2009 NHTS data | Upper bound: +14% | New demand from underserved travelers including elderly, young age, and travel-restricted with medical condition |
| Wadud et al. (2016) [5] | Literature-driven elasticity of VMT | +4% to +60% | Reduced generalized cost of driving |
| Stephens et al. (2016) [4] | Assumption based on multiplicative factors for travel demand | +20% to +160% | Easier travel due to traffic flow, crash avoidance, reduced cost of driving |
| Zhang et al. (2018) [47] | Activity-based model of Atlanta, GA area | +30% (per reduced vehicle) | Unoccupied relocation of private CAVs for meeting travel needs of household with reduced vehicle ownership |
| Harb et al. (2018) [48] | Naturalistic experiment, survey, and interview when providing chauffeur as a proxy for CAVs | +4% to +341% with central estimates of 83% increase | Travel pattern shift, longer and more frequent travels, unoccupied VMT (for a small sample size) |
| **This Study** | Estimation of VMT elasticity with respect to fuel- and time-inclusive marginal price of private vehicle driving using 2017 NHTS data | +2% to 47% | Reduced marginal cost of driving and heterogeneous response of different income groups (purpose: forecasting energy consumption impacts) |

# 6.    Conclusion

The aim of this study is to shed light on the possible travel and energy impacts of CAVs. To that end, we use microeconomic modeling, applied econometric techniques, and the most recent data available on household travel behavior to estimate average travel demand elasticities with respect to the price of fuel and travel time. We then leverage these elasticity estimates in a forecast of CAV-induced travel under a range of different realized changes to fuel economy and per-mile time costs.

We estimate an average elasticity of VMT demand with respect to the combined, fuel- and time-inclusive price per mile of -0.4. Allowing for heterogeneity in VMT elasticity by price channel (fuel vs. time) and income, we find that demand response to price increases is larger through the time channel (with an elasticity of -0.4) than through the fuel channel (with an elasticity of -0.1). We also find that richer households are more sensitive to the overall price of travel as well as the time cost.

Applying these fuel and time cost elasticities in our forecasting exercise, we find a large range of possible travel and energy impacts of CAV diffusion. A number of plausible scenarios for fuel economy and time cost changes are characterized by backfire, or a net rise in energy use. Backfire is more likely in higher income quantiles, where relatively less of a time cost reduction is required to offset the energy savings from fuel economy improvements. On average, a 38% reduction in time cost fully offsets a 20% fuel economy improvement enabled by CAVs.

Our results strongly suggest that travel demand will rise as a behavioral response to the diffusion of CAVs. Some of this rise will come from shifts away from other transportation modes, including public transit, cycling, and walking. Some will come from additional travel – such as new passenger trips, empty trips in between passenger travel, travel pattern change, breaking of pooled trips into several lower occupancy trips, and longer and more frequent trips necessitated by shifting home locations to peripheral zones. Regardless, this induced travel will pose a stiff challenge to policy goals for reductions in energy use, traffic congestion, and local and global air pollution.

The proper government response to CAV market penetration is not obvious. There is no "silver bullet" that can achieve all goals efficiently and equitably, and policies aimed at meeting some of these goals may make it more difficult to meet others. For instance, while it is natural to view our results as evidence that even greater fuel efficiency is needed, our study also underscores the limitations of vehicle energy efficiency improvements: they provide incentive to drive more, which offsets some environmental benefits and increases congestion. Taxation – another commonly cited policy tool for internalizing the negative externalities of driving – is also imperfect. Taxes are viewed by many as a more economically efficient policy instrument, but they are also sometimes viewed as regressive, because poorer households generally devote a greater proportion of their total budget to energy than richer ones. Vehicle connectivity may, on the one hand, actually enhance the cost-effectiveness of taxation in the transportation sector by offering the potential to tax VMT instead of (or in addition to) to fuel use.[20] On the other hand, the fact that wealthier households have more elastic demand than poorer ones in our context increases the risk of regressive welfare impacts of taxation.[21] Above all, policymakers should prioritize incentives for high-occupancy pooling, ride-sharing, and minimizing empty trips, as these have the potential for large reductions in fuel use at low cost to well-being.

Our analysis expresses induced travel and rebound in percentage terms, but it is instructive to consider the absolute magnitude of prospective changes in travel and energy due to CAVs. For instance, an assumed 15% average improvement in fuel economy is expected to save 10.56 billion gallons of gasoline equivalent (GGE) annually (26.4 billion USD), from a current consumption level of 88.85 billion GGE in light-duty vehicles. However, that number should be viewed as a best-case scenario. CAVs with the same 15% fuel economy advantage would very likely induce travel that would offset some of those savings. Based on our estimate, at 100% market penetration, CAVs may result in anywhere between the aforementioned 10.56 billion GGE annual decrease and a 15.26 billion GGE (17.2%, or 38.15 billion USD) annual *increase*.

---

[20] This is seen as desirable because, while fuel use is highly correlated with greenhouse gas emissions, it is much more weakly correlated with local air pollution, congestion, and accident risk (see, e.g., [55]).

[21] The relationship between demand elasticity and income is an important input into distributional welfare analysis; see [56,57]).

While the present study uses U.S. data to quantify the energy rebound caused by CAV penetration, the methodology that we develop here is general and can be applied to other regions of the world, where travel is less heavily reliant on private vehicles. Future research should also aim to compare the broader social benefits of CAV travel with their social costs, considering the value and frequency of driving and all the externalities that it produces. Finally, there remains a large degree of uncertainty in the attributes, costs, and benefits of connected and automated vehicles, which in turn makes it difficult to forecast and react to future travel and energy behaviors. Even at this early stage of CAV technology maturity, however, it is vital to consider the potential of CAVs to induce significant new travel and energy use.

**Data Availability:**

All data utilized within this study are publicly available. The National Household Travel Survey is available through the U.S. Federal Highway Administration – Department of Transportation (https://nhts.ornl.gov/). Information regarding the procedures, survey methodology, and data processing can be found in the 2017 NHTS User Guide [25]. Fuel economy testing data is available through the U.S. Environmental Protection Agency's National Vehicle and Fuel Emissions Laboratory in Ann Arbor, Michigan (https://www.fueleconomy.gov/feg/download.shtml).

**Acknowledgment:**

M.T. acknowledges the support of Dow Sustainability Fellows Program at the University of Michigan Graham Institute. The authors thank Dr. Zia Wadud of Centre for Integrated Energy Research and Institute for Transport Studies at University of Leeds, UK for helpful discussion and comments on the elasticity modeling approach. M.T. thanks Mr. Ali Rafei of Institute for Social Research at University of Michigan for assistance in understanding the NHTS dataset.

# Appendix

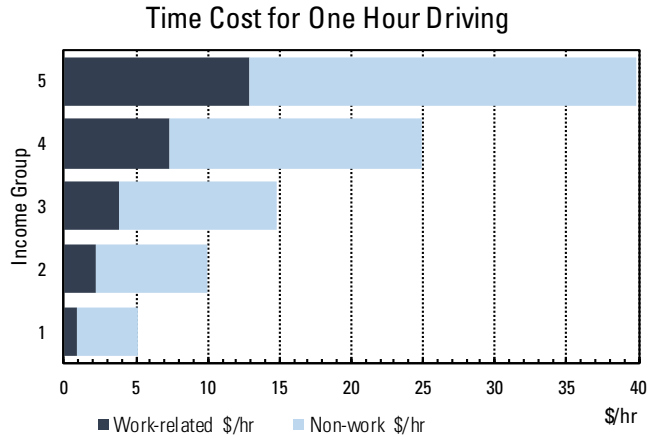

Figure A1. Bars denote the travel time cost of one hour of driving, for the average household in each income group. The national average is 19.56 $/hr.

Table A1. Estimation result of 2009 NHTS

| Income Group | 1st Income Group | 2nd Income Group | 3rd Income Group | 4th Income Group | 5th Income Group | U.S. Average | % difference of average with average of 2017 NHTS |
|---|---|---|---|---|---|---|---|
| | | | Panel A: **Model 3** | | | | |
| $\hat{\varepsilon}_f$ | -0.161*** (0.027) | -0.119*** (0.014) | -0.101*** (0.016) | -0.137*** (0.020) | -0.140*** (0.022) | -0.128*** (0.022) | 29.4% |
| $\hat{\varepsilon}_t$ | -0.353*** (0.055) | -0.444*** (0.049) | -0.498*** (0.051) | -0.518*** (0.039) | -0.552*** (-0.051) | -0.501*** (0.055) | 25.1% |
| | | | Panel B: **Model 4** | | | | |
| $\hat{\varepsilon}_{vmt}$ | -0.291*** (0.050) | -0.394*** (0.048) | -0.459*** (0.037) | -0.488*** (0.049) | -0.513*** (0.054) | -0.451*** (0.051) | 15.0% |

Dependent variable is $\log(VMT)$. Asterisks denote 1 (***), 5 (**), and 10 (*) percent significance levels, based on p-value. Clustered standard errors are reported in parentheses. Regressions include all controls and fixed effects described in the main text. Standard errors are clustered by MSA, and observations are weighted by household sampling weights. The dollar value is unadjusted between 2017 and 2009. The sample size for both models is 134,482. The pseudo $R^2$ of the regression is 0.213 in Panel A and 0.198 in Panel B.

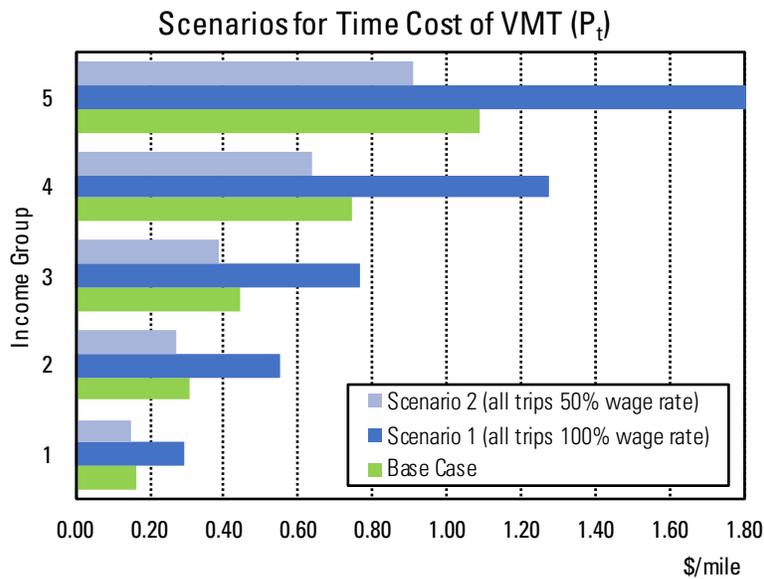

Figure A2. Scenarios designed for different definitions of TTC. 'Base Case' assigns 100% hourly wage to work trips and 50% hourly wage to non-work trips. 'Scenario 1' assigns 100% hourly wage to all trips while 'Scenario 2' assigns 50% hourly wage to all trips.

Table A2. Results of robustness check with respect to the definition of TTC

| Income Group | 1st Income Group | 2nd Income Group | 3rd Income Group | 4th Income Group | 5th Income Group | U.S. Average |
|---|---|---|---|---|---|---|
| | | | Panel A: **Scenario 1** | | | |
| $\hat{\varepsilon}_{vmt}$ (Model 4) | -0.140*** (0.026) | -0.197*** (0.029) | -0.230*** (0.028) | -0.261*** (0.022) | -0.251*** (0.028) | -0.225*** (0.028) |
| $\hat{\varepsilon}_t$ (Model 3) | -0.159*** (0.034) | -0.226*** (0.031) | -0.256*** (0.028) | -0.272*** (0.022) | -0.283*** (0.029) | -0.229*** (0.027) |
| | | | Panel B: **Scenario 2** | | | |
| $\hat{\varepsilon}_{vmt}$ (Model 4) | -0.283*** (0.053) | -0.373*** (0.055) | -0.432*** (0.051) | -0.481*** (0.040) | -0.460*** (0.046) | -0.422*** (0.053) |
| $\hat{\varepsilon}_t$ (Model 3) | -0.318*** (0.069) | -0.552*** (0.062) | -0.511*** (0.056) | -0.543*** (0.045) | -0.566*** (0.057) | -0.459*** (0.055) |

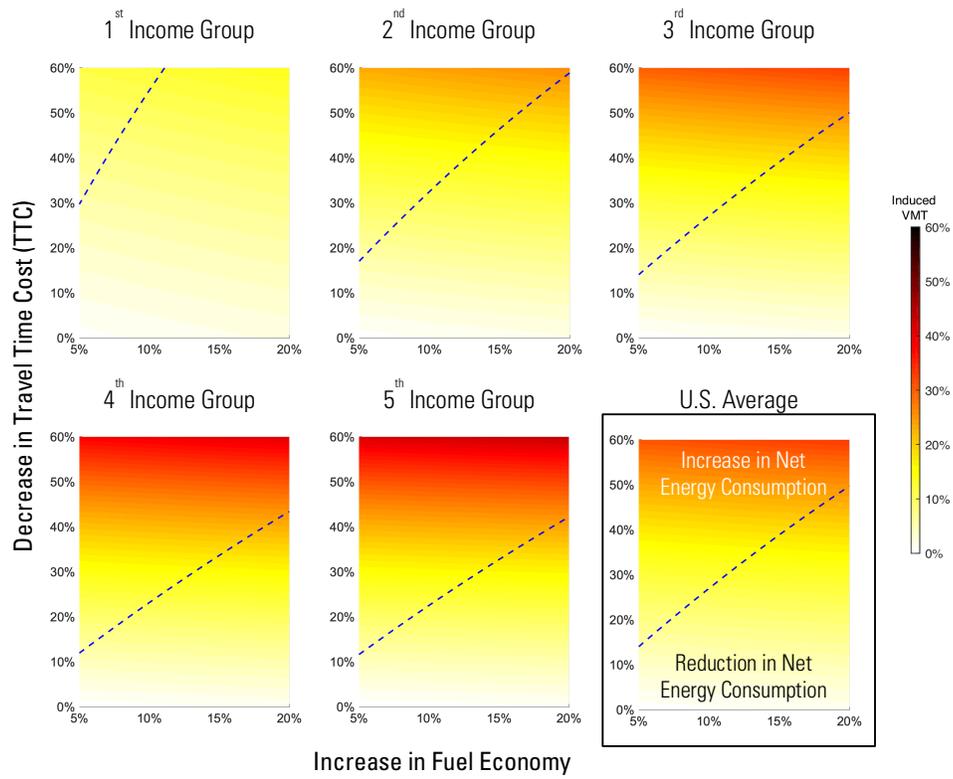

Figure A3. Heat maps of induced travel using Model 4. All points above dashed curves are characterized by backfire in net energy consumption.